\documentclass{amsart}
\usepackage[margin=1.15in]{geometry}

\newcommand{\NN}{{\mathbb{N}}}
\newcommand{\PP}{{\mathbb{P}}}
\newcommand{\Reals}{{\mathbb{R}}}
\newcommand{\Cmplx}{{\mathbb{C}}}
\newcommand{\EE}{{\mathbb{E}}}
\newcommand{\Ints}{{\mathbb{Z}}}
\newtheorem{thm}{Theorem}[section]

\newtheorem{cor}[thm]{Corollary}
\newtheorem{lem}[thm]{Lemma}
\theoremstyle{definition}

\begin{document}

\title{A note on fractional moments for the one-dimensional continuum Anderson model}

\author{Eman Hamza, Robert Sims and G\"unter Stolz}

\begin{abstract}

We give a proof of dynamical localization in the form of exponential
decay of spatial correlations in the time evolution for the
one-dimensional continuum Anderson model via the fractional moments
method. This follows via exponential decay of fractional moments of
the Green function, which is shown to hold at arbitrary energy and
for any single-site distribution with bounded, compactly supported
density.

\end{abstract}

\date{\today}

\maketitle


%
%
%

\section{Introduction} \label{sec:intro}

The fractional moment method (FMM) was initially developed for the
discrete Anderson model in \cite{AM}. It has recently been extended
in \cite{AENSS} and \cite{BNSS} to cover continuum Anderson models,
where it was shown that, in any dimension $d\ge 1$, exponential
decay of fractional moments of the Green function, e.g.\
(\ref{eq:expodec}) below, implies dynamical and spectral
localization. In fact, as discussed below, the result on dynamical
localization which is obtained via the FMM is stronger than what is
obtained by other methods. The fractional moment condition
(\ref{eq:expodec}) has also been found to be a technically useful
tool in other contexts, for example in the proof of Poisson
statistics of eigenvalues of the Anderson model in finite volume
\cite{Minami} or vanishing of the d.\ c.\ electrical conductivity of
an electron gas \cite{AG}.

The main goal of this note is to fill a gap in the literature, which
is to show that the FMM applies to one-dimensional continuum
Anderson models. While localization properties of the
one-dimensional Anderson model are well understood via other
methods, given the mentioned applications it is useful to know that
a proof via fractional moments can be given. In dimension $d=1$
localization should hold in the Anderson model at all energies,
independent of the disorder strength. To conclude this via the FMM,
exponential decay of the fractional moments needs to be verified at
all energies. For the discrete Anderson model this was done in the
Appendix of \cite{Minami}.

Here we will do this for the continuum one-dimensional Anderson
model, which is a random operator in $L^2(\Reals)$ of the form
\begin{equation} \label{eq:Anderson}
H = H(\omega) = -\frac{d^2}{dx^2} + W + V_{\omega}.
\end{equation}
The background potential, $W$, is bounded, real-valued and
1-periodic, i.e. $W(x+1)=W(x)$. The random potential is given by
\begin{equation} \label{eq:randompot}
V_{\omega} = \sum_{n\in\Ints} \eta_n(\omega) f_n,
\end{equation}
where we will assume that the single site potentials $f_n$ are
translates $f_n(x)=f(x-n)$ of a non-negative and bounded function
$f$. Moreover, we suppose that $f$ is supported on $[0,1]$, and that
it is strictly positive on a non-trivial subinterval $J$ of $[0,1]$,
i.e.\ there exist constants $C\ge c>0$ such that
\begin{equation} \label{eq:singlesite}
c\chi_J \le f \le C\chi_{[0,1]}.
\end{equation}
For the random variables $\eta_n$, we assume that they are
independent and identically distributed. We will also assume that
their common distribution $\mu(A) = \PP(\eta_n\in A)$ has a bounded
density $\rho$ with compact support, i.e.\
\begin{equation} \label{eq:density}
\|\rho\|_{\infty} <\infty, \quad {\rm supp}( \rho) \subset [
\eta_{\rm min}, \eta_{\rm max}].
\end{equation}

Given any bounded interval $\Lambda$, we will denote by
$H_{\Lambda}=H_{\Lambda}(\omega)$ the restriction of $H$ to
$L^2(\Lambda)$ with Dirichlet boundary conditions. By
$G_{\Lambda}(z)=(H_{\Lambda}-z)^{-1}$ we denote the resolvent of
$H_{\Lambda}$. We write $\chi_x$ for the characteristic function of
the interval $[x,x+1]$. By $\|\cdot\|_2$ we will denote
Hilbert-Schmidt norm.

Our main result is
\begin{thm} \label{thm:1dfrac}
For any $E_0 \in \Reals$ there exists a number $s_0 \in (0,1)$ such
that for all $0<s \le s_0$ there are $\eta>0$ and $C<\infty$ such
that
\begin{equation} \label{eq:expodec}
\mathbb{E} \left(  \left\| \chi_x G_{\Lambda} (E) \chi_y
\right\|_2^s \right) \, \leq \, C \, e^{ - \eta |x-y|},
\end{equation}
holds for every interval $\Lambda$ with integer endpoints, all
integers $x, y \in \Lambda $ and $E \in (-\infty,E_0]$.
\end{thm}

Theorem~\ref{thm:1dfrac} will be proven in
Section~\ref{sec:fracmom}. As a preparation we will show in
Section~\ref{sec:Furstenberg} that for the continuum Anderson model
given by (\ref{eq:Anderson}) and (\ref{eq:randompot}) Furstenberg's
Theorem applies at all energies and thus, in particular, the
Lyapunov exponent is positive at all energies. We show this under
the weaker assumption that the distribution of the random coupling
constants $\eta_n$ has non-discrete support by combining results of
\cite{KS} and \cite{DSS1}.

Theorem~\ref{thm:1dfrac} implies dynamical and spectral localization
at all energies:

\begin{thm} \label{thm:localization}
For any $E_0 \in \Reals$ there exist $\eta>0$ and $C< \infty$ such
that
\begin{equation} \label{eq:dynloc}
\EE(\sup \|\chi_x g(H) P_{E_0}(H) \chi_y\|) \le Ce^{-\mu|x-y|}
\end{equation}
for all integers $x$ and $y$. Here the supremum is taken over all
Borel measurable functions $g$ which satisfy $|g|\le 1$ pointwise
and $P_{E_0}(H)$ is the spectral projection for $H$ onto
$(-\infty,E_0]$.

Also, $H$ almost surely has pure point spectrum with exponentially
decaying eigenfunctions.
\end{thm}

An argument which shows that Theorem~\ref{thm:localization} follows
from Theorem~\ref{thm:1dfrac} was provided in \cite{AENSS}. However,
to allow single-site potentials of small support as in
(\ref{eq:singlesite}) the proof in \cite{AENSS} needs to be slightly
modified. We indicate the changes at the end of
Section~\ref{sec:fracmom}

The particular choice $g(x)=e^{itx}$, $t\in \Reals$ arbitrary, shows
that (\ref{eq:dynloc}) is a result on dynamical localization. The
exponential decay bound on the right hand side is stronger than what
has been obtained with other methods. Note, however, that for the
discrete one-dimensional Anderson model the analog of
(\ref{eq:dynloc}) was already obtained in \cite{Kunz/Souillard} by a
method which has not yet been extended to the continuum, however,
see \cite{DS}. Spectral localization for $H$ is, of course, not new,
see e.g.\ \cite{DSS1} for a more general result. We include it here
for completeness and because it was shown in \cite{AENSS} how it
follows by an argument using the RAGE theorem from dynamical
localization and thus, via Theorem~\ref{thm:localization}, is a
consequence of (\ref{eq:expodec}).

As mentioned above, the discrete analog of our main result is proven
in an appendix of \cite{Minami}. For completeness, we include an
alternate proof of this fact in Section~\ref{sec:discrete}, where we
use methods similar to the ones in our proof of
Theorem~\ref{thm:1dfrac}. There we will also include a new proof of
boundedness of the fractional moments of Green's function for the
discrete Anderson model. For the "off-diagonal case", $x\not= y$ in
(\ref{eq:expodecdis}), this slightly streamlines earlier arguments,
e.g.\ \cite{AM,Graf}, by using a change-of-variables argument which
was developed for the continuum FMM in \cite{AENSS}. A similar
strategy was used in the context of {\it unitary Anderson models} in
\cite{HJS}.

Our proof of Theorem~\ref{thm:1dfrac} in Section~\ref{sec:fracmom}
uses Pr\"ufer variables which require to work at real energy $E$.
The finite volume resolvent $G_{\Lambda}(E)$ is almost surely well
defined as $H_{\Lambda}$ has discrete eigenvalues which are strictly
monotone in all the random parameters. For some applications and to
also have a result for infinite volume it is of interest to be able
to extend our main result to complex energy, i.e.\ to consider
energies $E+i\varepsilon$ in Theorem~\ref{thm:1dfrac} and its
discrete analog Theorem~\ref{thm:1dfracdis} with bounds which are
uniform in $\varepsilon>0$. As discussed in
Section~\ref{sec:remarks}, this can easily be done for
Theorem~\ref{thm:1dfracdis}. While we expect the same to hold for
the continuum, it does not seem to follow with our method of proof.

In order to make our presentation self-contained, we will provide a
variety of facts, well-known to those familiar with a-priori
solution bounds and the Pr\"ufer formalism, in an Appendix.

\section{Furstenberg at all energies} \label{sec:Furstenberg}

In this section we consider the continuous one-dimensional Anderson
model defined by (\ref{eq:Anderson}) and (\ref{eq:randompot}) under
the weaker assumption that the coupling constants have non-discrete
distribution, i.e.\
\begin{equation} \label{eq:nondis}
\mbox{supp}\, \mu \quad \mbox{is not discrete}.
\end{equation}

For fixed $E\in \Reals$, let $T(\eta,E)$ be the transfer matrices of
$-u''+Wu+\eta fu=Eu$ from 0 to 1 and $G(E)$ the Furstenberg group to
energy $E$, i.e. the closed subgroup of $SL(2,\Reals)$ generated by
the matrices $T(\eta,E)$ with $\eta$ varying in the support of the
single site distribution $\mu$.

The goal of this section is to prove the following result, which is
optimal with respect to the use of assumption \eqref{eq:nondis} and
thus of some interest by itself.
\begin{thm} \label{thm:Furst}
For the continuum one-dimensional Anderson model given by
(\ref{eq:Anderson}), (\ref{eq:randompot}) and (\ref{eq:nondis}), the
Furstenberg group $G(E)$ is non-compact and strongly irreducible for
all $E \in \Reals$.
\end{thm}

For the definition of strong irreducibility see \cite{BL}. By
Furstenberg's Theorem \cite{BL}, the above result implies that the
Lyapunov exponent associated with $G(E)$ is positive for all
energies $E\in \Reals$. That $\mu$ has non-discrete support is
crucial here. Examples have been constructed showing that
non-trivial but discretely supported single site distributions can
lead to a discrete set of critical energies where $G(E)$ is compact
or not strongly irreducible (and the Lyapunov exponent may vanish),
see \cite{DSS2} or Section~5 of \cite{DLS}.

Theorem~\ref{thm:Furst} follows from applying a slight
generalization of the main result in \cite{KS}, see
Theorem~\ref{thm:KS} below, to the methods developed in \cite{DSS1}.
For the sake of completeness, we outline this argument.

We begin by stating a generalization of the result in \cite{KS}. Let
$Q: \Reals \to \Reals$ be locally integrable and for $j=0,1$, take
$u_j: \mathbb{R} \to \mathbb{C}$  to be solutions of
\begin{equation}
-u_j'' + Qu_j = 0,
\end{equation}
neither of which are identically zero. For any $V:\mathbb{R} \to
\mathbb{R}$ with $V \in L^1( \Reals)$ and support contained in
$[0,1]$, denote by $u_{( \lambda)}$ the solution of
\begin{equation}
-u'' + (Q+ \lambda V)u = 0
\end{equation}
which satisfies $u_{(\lambda)}(x) = u_0(x)$ for all $x<0$. Here we
may consider coupling constants $\lambda \in \mathbb{C}$. The
question of interest in this context is: Given a non-trivial
function $V$, for how many values of $\lambda$ is it possible that
the solution $u_{(\lambda)}$, which for $x<0$ coincides with $u_0$,
is proportional to $u_1$ for $x>1$? The case where $u_0 = u_1$ is
discussed in \cite{KS}. Following their arguments, we define the
Wronskian
\begin{equation} \label{eq:defb}
b( \lambda) = W \left[ u_1, u_{(\lambda)} \right](x) =
u_1(x)u_{(\lambda)}'(x) - u_1'(x)u_{(\lambda)}(x)
\end{equation}
for $x >1$. The $\lambda$-set in question is given by the zeros of
$b$.
\begin{thm} \label{thm:KS}
If $V$ is not identically zero and either
\begin{equation} \label{eq:equal}
\mbox{$u_0 = u_1$ (and possibly complex-valued)}
\end{equation}
or
\begin{equation} \label{eq:real}
\mbox{$u_0$ and $u_1$ are real-valued},
\end{equation}
then the zeros of $b$ form a discrete set.
\end{thm}

In \cite{KS} this result is stated and proven for the case
$u_0=u_1$. However, for the case of real-valued solutions $u_0$ and
$u_1$, the proof provided in \cite{KS} goes through without change
if $u_0\not= u_1$. We will use both versions of this result below.

\vspace{.3cm}

\begin{proof}(of Theorem~\ref{thm:Furst})
Fix $E_0 \in \mathbb{R}$. Let $D(E) = \mbox{Tr}\left[ T(0,E)
\right]$ denote the discriminant of $- d^2/dx^2+ W$. The first step
in our proof demonstrates that, without loss of generality, we may
assume both $0 \in \mbox{supp}( \mu)$ and $D(E_0) \notin \{ -2, 0, 2
\}$. This is easily seen by adjusting the periodic background
$V_{\rm per}$. In fact, let $\eta_0$ be an accumulation point for
$\mbox{supp}( \mu)$. Consider $\tilde{D}(E) = \mbox{Tr} \left[ T(
\eta_0, E) \right]$, the discriminant of $- \frac{d^2}{dx^2} +
\tilde{W}$ where
\begin{equation} \label{eq:shiftpot}
\tilde{W} = W + \eta_0 \sum_{n \in \mathbb{Z}} f( \cdot - n) \, .
\end{equation}
Clearly
\begin{equation}
H_{\omega} = \tilde{W} +  \sum_{n \in \mathbb{Z}} \tilde{\eta}_n(
\omega) f( \cdot - n) \, ,
\end{equation}
where the random variables $\{ \tilde{\eta}_n \}$ have distribution
$\tilde{\mu}$ defined by $\tilde{\mu}(M) = \mu(M + \eta_0)$, i.e. $0
\in \mbox{supp}( \tilde{\mu})$. If $\tilde{D}(E_0) \notin \{ -2, 0,
2\}$, then we have completed the first step of this proof. If
$\tilde{D}(E_0) \in \{ -2, 0, 2\}$, then $E_0$ is an eigenvalue of
an operator with quasi-periodic boundary conditions. To see this,
define the family of self-adjoint operators
\begin{equation}
H_{\lambda, \theta} = - \frac{d^2}{dx^2} + \tilde{W} + \lambda f
\quad \mbox{on } [0,1]
\end{equation}
with boundary conditions $u(1) = e^{i \theta}u(0)$ and $u'(1) = e^{i
\theta}u'(0)$. It is clear that $E$ is an eigenvalue of $H_{\lambda,
\theta}$ if and only if the corresponding discriminant
$\mbox{Tr}\left[ T(\eta_0+\lambda,E) \right]$ is $2 \cos( \theta)$.
We conclude that if $\tilde{D}(E_0)= \mbox{Tr}\left[ T(\eta_0,E_0)
\right]\in \{ -2, 0, 2\}$, then $E_0$ is an eigenvalue of $H_{0,
\pi}$, $H_{0, \frac{\pi}{2}}$, or $H_{0, 0}$ respectively. Since $f
\geq 0$ and $f \neq 0$, analytic perturbation theory, see e.g.\
\cite{Kato}, implies that there exists $\delta >0$ such that for all
$\lambda \in (- \delta, \delta) \setminus \{ 0 \}$, $E_0$ is not an
eigenvalue of
 $H_{\lambda, \pi}$, $H_{\lambda, \frac{\pi}{2}}$, and $H_{\lambda, 0}$. This uses that all the eigenvalues of $H_{\lambda,\theta}$ are analytic and strictly increasing in $\lambda$, the latter being due to the Feynman-Hellmann formula which shows that (\ref{eq:singlesite}) suffices to get positivity of the $\lambda$-derivative of eigenvalues.

 As $\eta_0$ was an accumulation point, there exists
 $\lambda_1 \in (-\delta, \delta) \setminus \{ 0 \}$ such that $\eta_1 = \eta_0 + \lambda_1 \in \mbox{supp}( \mu)$.
 Defining $\tilde{\tilde{W}}$ analogously to (\ref{eq:shiftpot}) with $\eta_0$ replaced by $\eta_1$, we
 have completed step 1.

 Step 2 of this proof demonstrates the validity of Theorem~\ref{thm:Furst} in the event that
 $D(E_0) \in (-2,2) \setminus \{0\}$, i.e. $E_0$ is in a band of $-d^2/dx^2+W$ without being at the ``band center''. Let $\phi_{\pm}$ denote the linearly independent
  Floquet solutions of $- \phi'' + W \phi = E_0 \phi$, see e.g.~\cite{DSS1} for details.
 Denote by $u_{(\eta)}$ the solution of
 \begin{equation}
 -u'' +(W + \eta f)u = E_0 u
 \end{equation}
which satisfies
\begin{equation}
u_{(\eta)}(x) = \left\{ \begin{array}{cc} \phi_+(x) & \mbox{for } x
< 0, \\ a( \eta) \phi_+(x) + b( \eta) \phi_-(x) & \mbox{for } x >1.
\end{array} \right.
\end{equation}
A simple Wronskian argument shows that $a( \eta) \neq 0$ for all
$\eta$, and by Theorem~\ref{thm:KS} (under condition
(\ref{eq:equal})), the set $\{ \eta \in \mathbb{C} : b( \eta) = 0
\}$ is discrete. Since the support of $\mu$ is not discrete, there
exists a $\eta_0 \in \mbox{supp}( \mu) \setminus \{0\}$ for which
$b( \eta_0) \neq 0$. It is shown in \cite{DSS1} that $G(E_0)$
contains a subgroup which is conjugate to the group generated by the
matrices
\begin{equation}
Q^{-1} \left( \begin{array}{cc} \rho & 0 \\ 0 & \overline{\rho}
\end{array} \right) Q \quad \mbox{and} \quad Q^{-1} \left(
\begin{array}{cc} a( \eta_0) & \overline{b(\eta_0)} \\ b(\eta_0) &
\overline{a(\eta_0)} \end{array} \right) Q \quad \mbox{where} \quad
Q = \frac{1}{2} \left( \begin{array}{cc} 1 & -i \\ 1 & i \end{array}
\right)  \, ,
\end{equation}
and the numbers $\rho$ and $\overline{\rho}$ are the Floquet
multipliers, i.e.\ the eigenvalues of the transfer matrix $T(0,
E_0)$. $D(E_0) \in (-2,2)\setminus \{0\}$ means that $\rho =
e^{i\omega}$ with $\omega \in (0,\pi)\setminus \{\pi/2\}$. Using
this and the explicit form of this group, it was shown to be
non-compact and strongly irreducible in \cite{DSS1}. The same
readily follows for $G(E_0)$.

Step 3 finishes the proof in the case that $|D(E_0)| >2$, i.e. $E_0$
is in a gap of $-d^2/dx^2+W$. In this case, there exist real-valued
linearly independent solutions $u_{\pm}$, each not identically zero,
of
\begin{equation}
-u'' +Wu = E_0 u
\end{equation}
with $u_{\pm}$ in $L^2$ near $\pm \infty$. Similar to above, we
denote by $u^{\pm}_{(\eta)}$ the solution of
 \begin{equation}
 -u'' +(W + \eta f)u = E_0 u
 \end{equation}
which satisfies
\begin{equation}
u^{\pm}_{(\eta)}(x) = \left\{ \begin{array}{cc} u_{\pm}(x) &
\mbox{for } x < 0, \\ a_{\pm}( \eta) u_{\pm}(x) + b_{\pm}( \eta)
u_{\mp}(x) & \mbox{for } x >1. \end{array} \right.
\end{equation}
Using Theorem~\ref{thm:KS} (under condition (\ref{eq:real})) for
each of the four pairs $(u^{\pm},u^{\pm})$, one finds that the set
\begin{equation}
\{ \eta \in \mathbb{C} : a_+( \eta) b_+( \eta)a_-( \eta) b_-(\eta) =
0 \}
\end{equation}
is discrete. Picking $\eta_0 \in \mbox{supp}(\mu) \setminus \{0 \}$
for which $a_+( \eta_0) b_+( \eta_0)a_-( \eta_0) b_-(\eta_0) \neq
0$, we will prove that the subgroup generated by $T(0,E_0)$ and $T(
\eta_0, E_0)$ is non-compact and strongly irreducible repeating
arguments from \cite{DSS1}.

Since $|D(E_0)| >2$, $T(0,E_0)$ has eigenvalues $\rho$ and
$\rho^{-1}$ with $\rho >1$ or $\rho <-1$. Denote by
\begin{equation}
v_{\pm} = \left( \begin{array}{c} u_{\mp}(0) \\ u_{\mp}'(0)
\end{array} \right)
\end{equation}
the eigenvectors of $T(0,E_0)$ corresponding to $\rho$ and
$\rho^{-1}$, respectively. Clearly, $w_n = T(0,E_0)^n v_+$ is
unbounded, and therefore, the subgroup generated by $T(0,E_0)$ alone
is non-compact. As we have shown that this group is non-compact, to
prove that it is also strongly irreducible, we need only show that
each direction is mapped onto at least three distinct directions by
this group, see e.g. \cite{BL}. First, suppose $v$ is not in the
direction of $v_+$ or $v_-$. Then, the sequence $w_n = T(0,E_0)^n v$
produces arbitrarily many directions (as $w_n$ approaches the stable
manifold generated by $v_-$). If $v$ is in the direction of $v_+$ or
$v_-$, then $T(\eta_0, E_0)v$ is not as $a_+( \eta_0) b_+(
\eta_0)a_-( \eta_0) b_-(\eta_0) \neq 0$. By our previous argument
then, $\tilde{w}_n = T(0,E_0)^nT( \eta_0,E_0) v$ produces
arbitrarily many directions. This completes step 3 and the proof of
Theorem~\ref{thm:Furst}.
\end{proof}

\section{Proof of Theorem~\ref{thm:1dfrac}} \label{sec:fracmom}

Non-compactness and strong irreducibility of the Furstenberg group
$G(E)$, if known for all energies in an interval, leads to
consequences which go beyond positivity of the Lyapunov exponents.
To state the result which we need, denote by $T(n,k,E) =
T_{\omega}(n,k,E)$ the transfer matrix of $H$ at energy $E$ from $k$
to $n$, i.e.\ the $2\times 2$-matrix such that
\[ T(n,k,E) \left( \begin{array}{c} u(k) \\ u'(k) \end{array} \right) = \left( \begin{array}{c} u(n) \\ u'(n) \end{array} \right) \]
for all solutions of $-u'' + (W+V_{\omega})u=Eu$.

\begin{lem} \label{lem:furstenberg}
Let $I \subset \Reals$ be a compact interval such that $G(E)$ is
non-compact and strongly irreducible for every $E\in I$. Then there
exist $\alpha_1>0$, $\delta>0$ and $n_0 \in \NN$ such that for all
$E\in I$, $n\ge n_0$ and $x \in \Reals^2$ normalized,
\[ \EE(\|T(n,0,E)x\|^{-\delta}) \le e^{-\alpha_1 n}. \]
\end{lem}

This is essentially Lemma~5.2 of \cite{DSS1}. While the latter is
stated in a more concrete setting, the above slightly abstracted
version is what one gets from the argument provided in \cite{DSS1}
to which we refer for the proof.

Thus, under the assumptions of Theorem~\ref{thm:1dfrac}, we conclude
from Theorem~\ref{thm:Furst} that Lemma~\ref{lem:furstenberg}
applies to {\it every} compact interval $I$. To prove
Theorem~\ref{thm:1dfrac} it suffices to consider energies $E \in I
:= [E_1, E_0]$, where $E_1$ is a deterministic and strict lower
bound of the potential $W+V_{\omega}$ (which exists by our
assumptions). For energies below $E_1$ exponential decay of the
right hand side of (\ref{eq:expodec}) is a deterministic consequence
of Combes-Thomas bounds, e.g.\ \cite{Stollmann}.

\vspace{.3cm}

Our main tools in reducing (\ref{eq:expodec}) to
Lemma~\ref{lem:furstenberg} are the Pr\"ufer amplitudes and phases
corresponding to solutions of $H_{\Lambda} u = Eu$. We introduce
these as follows. Write $\Lambda = [a,b]$ for integers $a$, $b$. For
any $E\in \Reals$, $c\in [a,b]$ and $\theta \in \Reals$ we denote by
$u_c(x,E,\theta)$ the solution of $-u''+(W+V_{\omega})u=Eu$ which
satisfies $u(c)=\sin \theta$ and $u'(c)= \cos \theta$. By regarding
this solution and its derivative in polar coordinates, we define the
Pr\"ufer amplitude, $R_c(x,E,\theta)$, and the Pr\"ufer phase,
$\phi_c(x,E,\theta),$ by writing
\begin{equation} \label{eq:pruferdef}
u_c(x,E,\theta)= R_c(x,E,\theta) \sin \phi_c(x,E,\theta) \quad
\mbox{and} \quad u_c'(x,E,\theta) = R_c(x,E,\theta) \cos
\phi_c(x,E,\theta).
\end{equation}
For fixed $E$, we declare $\phi_c(c,E,\theta)=\theta$ and require
continuity of $\phi$ in $x$. In this manner we define uniquely the
functions $R_c(x,E,\theta)$ and $\phi_c(x,E,\theta)$ which are
jointly continuous in $x$ and $E$.

For the remainder of this section, finite positive constants which
can be chosen uniform in the given context may change their value
from line to line.

\begin{proof}(of Theorem~\ref{thm:1dfrac})
We may assume that the integers $x$, $y$ satisfy $x\le y$ (if $x>y$
use that $\|\chi_x G_{\Lambda}(E)\chi_y\|_2 = \|(\chi_x
G_{\Lambda}(E)\chi_y)^*\|_2 = \|\chi_y G_{\Lambda}(E)\chi_x\|_2$).
Since $H_{\Lambda}$ satisfies Dirichlet boundary conditions at both
$a$ and $b$, the Green's function can be written in terms of the
solutions $u_a = u_a(\cdot, E, 0)$ and $u_b = u_b(\cdot,E,0)$ if $E$
is not in the spectrum of $H_{\Lambda}$. In this case
\begin{equation} \label{eq:green}
G_{\Lambda}(s,t;E) = \frac{1}{W(u_a,u_b)} \left\{ \begin{array}{ll}
u_a(s) u_b(t) & \mbox{if $s\le t$}, \\ u_a(t) u_b(s) & \mbox{if
$s>t$.} \end{array} \right.
\end{equation}
where $W(u_a,u_b) = u_a u_b'-u_a' u_b$ is the Wronskian of the
solutions $u_a$ and $u_b$. Let us first consider the case $x<y$. As
explained in Section~\ref{sec:intro}, a fixed $E$ is almost surely
in the resolvent set of $H_{\Lambda}$, and hence, for almost every
$\omega$, we have that
\begin{eqnarray} \label{eq:greenbound}
\| \chi_x G_{\Lambda}(E) \chi_y \|_2^2 &= & \int_x^{x+1} \int_y^{y+1} \Big|\frac{u_a(s)u_b(t)}{W(u_a ,u_b )}\Big|^2 \, dt \, ds  \\
& \leq & \frac{1}{|W(u_a,u_b)|^2}\int_x^{x+1} \int_y^{y+1} |R_a(s,E,0)R_b(t,E,0)|^2 \, dt \, ds \nonumber \\
&\leq & \frac{C}{|W(u_a,u_b)|^2}|R_a(x, E,0)R_b(y, E,0)|^2.
\nonumber
\end{eqnarray}
Here (\ref{eq:solbd}) in Lemma~\ref{lem:solest} in the Appendix was
used, where a uniform constant can be chosen since $W+V_{\omega}-E$
has local $L^1$-bounds which can be chosen uniformly in $\omega$ and
$E\in I$. If $x=y$, then the representation (\ref{eq:green}) leads
to two terms in (\ref{eq:greenbound}), but Lemma~\ref{lem:solest}
leads to the same resulting bound. Therefore, we have that
\begin{eqnarray}
\mathbb{E} \left( \| \chi_x G_{\Lambda}(E) \chi_y \|_2^s \right)  &
\leq & C\mathbb{E} \left(  \frac{R^s_a(x, E,0)R^s_b(y,
E,0)}{|W(u_a,u_b)|^s}
 \right) \\
 &= & C \widehat{\mathbb{E}}\left(\int_{\eta_{\rm min}}^{\eta_{\text{max}}}  \frac{R^s_a(x, E,0)R^s_b(y,
E,0)}{|u_a'(x)u_b(x)-u_a(x)u_b'(x)|^s}\rho(\eta_x)d\eta_x\right),
\nonumber
\end{eqnarray}
where $\widehat{\mathbb{E}}$ denotes the expectation with respect to
the random variables $\{\eta_n\}_{n\in\Ints\backslash\{x\}}$.

By construction, the random variable $\eta_x$ multiplies the single
site with support on $[x,x+1]$, and therefore both $R^s_a(x, E,0)$
and $R^s_b(y, E,0)$ are independent of $\eta_x$. {F}rom this, we
conclude that
\[
\mathbb{E} \left( \| \chi_x G_{\Lambda}(E) \chi_y \|_2^s \right) \,
\leq
C\widehat{\mathbb{E}}\left(\frac{R^s_b(y,E,0)}{R_b^s(x,E,0)}\int_{\eta_{\rm
min}}^{\eta_{\rm max}}
\frac{\rho(\eta_x)}{\left|\sin(\phi_b(x,E,0)-\phi_a(x,E,0))\right|^s}d\eta_x\right).
\]
The inner integral above may be bounded using Lemma~\ref{lem:1dbd}
which is proven below. Using this result, we find that
\begin{equation} \label{eq:baseintbd}
\mathbb{E} \left( \| \chi_x G_{\Lambda}(E) \chi_y \|_2^s \right) \,
\leq
C\widehat{\mathbb{E}}\left(\frac{R^s_b(y,E,0)}{R_b^s(x,E,0)}\right).
\end{equation}
It follows from the definition of Pr\"ufer variables that
\[
R_b^2(x,E,0) \, = \, R_b^2(y,E,0) \, R_y^2(x,E, \phi_b(y,E,0)),
\]
and therefore, the right hand side of (\ref{eq:baseintbd}) can be
written in terms of the product of transfer matrices
\begin{eqnarray} \label{eq:prod}
\frac{R_b^s(y,E,0)}{R_b^s(x,E,0)} & = & \frac{1}{R_y^s(x,E, \phi_b(y,E,0))} \\
& = & \left\|T(x,y, E) \left( \begin{array}{c} \sin \phi_b(y,E,0)\\
\cos \phi_b(y,E,0) \end{array} \right) \right\|^{-s} \,. \nonumber
\end{eqnarray}
$T(x,y,E)$ depends on the random variables $\eta_x$, \ldots,
$\eta_{y-1}$, while $\phi_b(y,E,0)$ depends on $\eta_y$,
$\eta_{y+1}$, \ldots. Thus Lemma~\ref{lem:furstenberg} (which holds
equally well for the ``backwards'' transfer matrices considered
here) can be applied to the right-hand-side of (\ref{eq:prod}),
yielding (\ref{eq:expodec}) as claimed.
\end{proof}

This completes the proof of Theorem~\ref{thm:1dfrac} given
Lemma~\ref{lem:1dbd}. We now state and prove this fact.

\begin{lem}\label{lem:1dbd}
For any bounded interval $I\subset\Reals$ and $0<s<1$, there exists
$C<\infty$, such that
\begin{equation} \label{eq:intbd}
\int_{\eta_{\rm min}}^{\eta_{\rm max}}
\frac{\rho(\eta_x)}{\left|\sin(\phi_b(x,E,0)-\phi_a(x,E,0))\right|^s}d\eta_x
\leq C
\end{equation}
for any integer interval $[a,b]$, any integer $x \in [a,b]$, and
$E\in I$.
\end{lem}

\begin{proof}
Observe that the random variable $\phi_a(x,E,0)$ is determined by
the parameters $\{ \eta_n \}_{n=a}^{x-1}$, whereas $\phi_b(x,E,0)$
depends on $\{ \eta_n \}_{n=x}^{b-1}$. This suggests the change of
variables $t(\eta_x)=\phi_b(x,E,0)$. The result of
Lemma~\ref{lem:lamderphi} says in the current context that
\[
t'(\eta_x) = \frac{1}{R_b^2(x,E,0)} \int_x^{x+1} f_x(t)
u_b^2(t,E,0)\,dt.
\]
Using the condition (\ref{eq:singlesite}) on the single site
potential in combination with Lemmas~\ref{lem:solest} and
\ref{lem:L2b} we find constants such that
\begin{equation} \label{eq:uplowbound}
C_1 R_b^2(x,E,0) \le \int_x^{x+1} f_x(t) u_b^2(t,E,0)\,dt \le C_2
R_b^2(x,E,0)
\end{equation}
and thus
\begin{equation} \label{eq:etaprime}
0< C_1 \le t'(\eta_x) \le C_2 < \infty
\end{equation}
uniformly in $\omega$ and $E\in I$. Therefore, we have that
\begin{equation} \label{eq:cov}
\int_{\eta_{\rm min}}^{\eta_{\rm max}}
\frac{\rho(\eta_x)}{\left|\sin(\phi_b(x,E,0)-\phi_a(x,E,0))\right|^s}d\eta_x
\leq C \| \rho \|_{\infty} \int_{t(\eta_{\rm min})}^{t(\eta_{\rm
max})} \frac{1}{\left|\sin(t -\phi_a(x,E,0))\right|^s}dt.
\end{equation}
But by (\ref{eq:etaprime}) we also have $|t(\eta_{\rm max})-t(
\eta_{\rm min})| \leq C$ uniformly in $\omega$ and $E\in I$. The
inequality claimed in (\ref{eq:intbd}) now follows using
(\ref{eq:cov}) and the fact that the resulting integrand has only a
finite number of integrable singularities in any bounded interval,
independent of the phase shift $\phi_a(x,E,0)$.
\end{proof}

\vspace{.3cm}

We end this section with some comments on the {\it proof of
Theorem~\ref{thm:localization}}, which follows by a slight
adaptation of the proof of Theorem~1.1 in \cite{AENSS}. Essentially,
this amounts to avoiding use of the covering condition for the
single site potential required in \cite{AENSS} and thus allowing for
single site potentials of small support as in (\ref{eq:singlesite}).

To prove (\ref{eq:dynloc}) for given $E_0\in \Reals$, we may again
work on the interval $I = [E_1,E_0]$ with $E_1$ as above. As in
Section~2 of \cite{AENSS} define, for a finite interval $\Lambda$
and integers $x$, $y$,
\begin{equation} \label{eq:totvar}
Y_{\Lambda}(I;x,y) := \sup \left\{ \|\chi_x f(H^{\Lambda})
\chi_y\|\,:\,f\in C_c(I), \|f\|_{\infty} \le 1\right\},
\end{equation}
where $C_c(I)$ are the continuous functions with compact support
inside $I$. Let $E_n$ and $\psi_n$ denote the eigenvalues and
corresponding orthonormal eigenfunctions of $H^{\Lambda}$ and
$P_{\psi_n}$ be the orthogonal projector onto $\psi_n$. Thus
$f(H^{\Lambda}) = \sum_{n:E_n\in I} f(E_n) P_{\psi_n}$ and
\begin{eqnarray} \label{eq:Ybound}
Y_{\Lambda}(I;x,y) & \le & \sum_{n:E_n\in I} \|\chi_x P_{\psi_n} \chi_y\| \\
& = & \sum_{n:E_n\in I} \|\chi_x \psi_n\| \|\chi_y \psi_n\|
\nonumber
\end{eqnarray}
As in (\ref{eq:uplowbound}), using Lemmas~\ref{lem:solest} and
\ref{lem:L2b}, we have
\[ \|f_y^{1/2} \psi_n \|^2 = \int_y^{y+1} f_y(t) \psi_n^2(t)\,dt \ge C_1 (|\psi_n(y)|^2 + |\psi_n'(y)|^2) \]
and
\[ \|\chi_y \psi_n\|^2 \le C_2 (|\psi_n(y)|^2 + |\psi_n'(y)|^2) \]
uniformly in $\Lambda$, $n$ and $\omega$. Thus $\|\chi_y \psi_n\|
\le C\|f_y^{1/2} \psi_n\|$ and (\ref{eq:Ybound}) gives
$Y_{\Lambda}(I;x,y) \le C Q_1(I;x,y)$, with the eigenfunction
correlator
\[ Q_1(I;x,y) := \sum_{n:E_n\in I} \|\chi_x \psi_n\| \|f_y^{1/2} \psi_n\|.\]
From here the proof is completed as in \cite{AENSS}, where no
additional use of the covering condition is made.



\section{The discrete case} \label{sec:discrete}

The one-dimensional discrete Anderson model $h = h(\omega)$ acts on
$l^2(\mathbb{Z})$ as
  \begin{equation} \label{eq:discreteAnderson}
(h u)(n)= -u(n+1)-u(n-1)+ \eta_n(\omega) u(n).
\end{equation}
As before, we assume that the random variables $(\eta_n)$ are
i.i.d.\ with density $\rho$ satisfying (\ref{eq:density}).
  For $a,b \in\mathbb{Z}$, $a<b$, we write $[a,b]:=\{a, a+1,..., b\}$, for convenience. The restriction of $h$ to $\ell^2([a,b])$ is denoted by $h_{[a,b]}$, the Green function by $G_{[a,b]}(x,y;z) := \langle e_x, (h_{[a,b]}-z)^{-1} e_y \rangle$.

The following result was first proven by Minami in an appendix of
\cite{Minami}. We include it here to supplement our main result
Theorem~\ref{thm:1dfrac} with its discrete analogue and to provide a
somewhat different self-contained proof.

\begin{thm} \label{thm:1dfracdis}
There exists a number $s_0 \in (0,1)$ such that for all $0<s \leq
s_0$, the bound
\begin{equation} \label{eq:expodecdis}
\mathbb{E} \left(  |  G_{[a,b]}(x,y;E) |^s \right) \, \leq \, C \,
e^{ - \eta |x-y|},
\end{equation}
holds for all $x, y \in [a,b] $ and $E \in \Reals$. Here the numbers
$C>0$ and $\eta>0$ depend on $s$, however, they may be chosen
independent of $[a,b]$.
\end{thm}

For $E$ outside the spectrum of $h_{\omega}$ exponential decay of
Green's function follows from deterministic Combes-Thomas bounds.
Thus it will suffice to show (\ref{eq:expodecdis}) for energies $E$
in, say, $I=[-3+\eta_{min}, 3+\eta_{max}]$.

We start by establishing a uniform a priori bound on the left hand
side of (\ref{eq:expodecdis}). This is well known ever since the
ground breaking work \cite{AM}, but we opt to include a somewhat
streamlined proof, using a more recent change of variables idea.

\begin{lem} \label{lem:bounddis} Let $s\in(0,1)$.
There exists a number $C<\infty$ such that
\begin{equation} \label{eq:bounddis}
\mathbb{E} \left( |  G_{[a,b]}(x,y;E)|^s \right) \, \leq \, C,
\end{equation}
for all integers $a<b$ and $x, y \in [a,b]$ and $E \in \Reals$.
\end{lem}

\begin{proof}
For $x, y \in [a,b]$, $x\not= y$, write $h= \hat{h} + \eta_x P_x
+\eta_y P_y$, where $P_x = \langle e_x,\cdot \rangle e_x$, $P_y=
\langle e_y, \cdot \rangle e_y$. Also writing $P=P_x +P_y$ we get,
using Krein's formula,
\begin{equation} \label{eq:krein}
G_{[a,b]}(x,y;E)=\left[A^{-1}+\left( \begin{array}{cc} \eta_x & 0 \\
0 & \eta_y \end{array} \right) \right]^{-1}(x,y),
\end{equation}
with the $2\times 2$-matrix $A = P(\hat{h}-E)^{-1}P$.

We introduce the change of variables $\alpha =
\frac{1}{2}(\eta_x+\eta_y)$, $\beta = \frac{1}{2}(\eta_x-\eta_y)$.
With the self adjoint matrices $A_{\beta} := A^{-1} +\beta \left(
\begin{array}{cc} 1 & 0 \\ 0 & -1 \end{array} \right)$, the right
hand side of \eqref{eq:krein} becomes $[A_{\beta}+\alpha
I]^{-1}(x,y)$.
 Therefore,
\begin{eqnarray} \label{eq:subrule}
\lefteqn{\int_{\eta_{\rm min}}^{\eta_{\text{max}}}\int_{\eta_{\rm
min}}^{\eta_{\text{max}}}|G_{[a,b]}(x,y;E)|^s
d\mu(\eta_x)d\mu(\eta_y)} \\ & \leq &
 2 ||\rho||_\infty^2 \int_{-(\eta_{\rm max}-\eta_{\rm min})/2}^{(\eta_{\rm max}-\eta_{\rm min})/2} \int_{\eta_{\rm min}}^{\eta_{\rm max}} \|[A_{\beta}+\alpha I]^{-1}\|^s \,
 d\alpha \,d\beta. \nonumber
\end{eqnarray}

A general fact, see e.g.\ Lemma~4.1 of \cite{HJS}, says that there
is a constant $C=C(s,\eta_{\rm max},\eta_{\rm min})$ such that
\begin{equation} \label{eq:fracint}
 \int_{\eta_{\rm min}}^{\eta_{\text{max}}}\left\|\left[B+\alpha
 \mathbb{I}\right]^{-1}\right\|^s d\alpha \leq C
\end{equation}
for all dissipative $2\times 2$-matrices $B$ (i.e.\ matrices with
Im$\,B\ge 0$). In \eqref{eq:subrule} we only need to use this for
self adjoint matrices to conclude the required bound for the case
$x\not= y$. The diagonal case $x=y$ is easier since no change of
variable is required and Krein's formula directly reduces the claim
to the elementary analogue of \eqref{eq:fracint} for $1\times
1$-matrices, i.e.\ numbers, see \eqref{eq:elembound} below.
\end{proof}

\begin{proof} (of Theorem~\ref{thm:1dfracdis})
Without loss of generality we assume that $x<y$, using the resolvent
identity we see that
\[
 G_{[a,b]}(x,y;E)= [1+G_{[a,b]}(x,x-1;E)]G_{[x,b]}(x,y;E).
\]
It suffices to prove the exponential decay of $\mathbb{E} \left(|
G_{[x,b]}(x,y;E) |^s \right)$ for $s\le s_1$. Using
Lemma~\ref{lem:bounddis} and H\"older's inequality it then follows
that \eqref{eq:expodecdis} holds for $s\le s_1/2$.

We have
\begin{equation} \label{eq:greenformula}
G_{[x,b]}(s,t;E) = \frac{1}{W(u_x,u_b)} \left\{ \begin{array}{ll}
u_x(s) u_b(t) & \mbox{if $s\le t$}, \\ u_x(t) u_b(s) & \mbox{if
$s>t$}. \end{array} \right.
\end{equation}
Here $u_x$ and $u_b$ are the solutions of $-u(n-1)-u(n+1)+\eta_n
u(n)=Eu(n)$ with $u_x(x-1)=0$, $u_x(x)=1$, $u_b(b)=1$, $u_b(b+1)=0$.
The constant Wronskian of $u_x$ and $u_b$ is given by
\[
W(u_x,u_b)(n) = u_x(n+1) u_b(n)-u_x(n) u_b(n+1).
\]
Evaluating the Wronskian at $n=x$ and denoting by $\widehat{\EE}$
the expectation conditioned on $\eta_x$, we obtain that
\[
\EE \left( |  G_{[x,b]}(x,y;E) |^s \right) \, =
\widehat{\EE}\left(\int_{\eta_{\rm min}}^{\eta_{\text{max}}}
\dfrac{|u_b(y)|^s}{|u_b(x+1)+(E-\eta_x)u_b(x)|^s}\rho(\eta_x)d\eta_x\right).
\]
 Now the main task is to show that
 \begin{equation}\label{eq:transferbd}
\int_{\eta_{\rm min}}^{\eta_{\rm max}}  \dfrac{|u_b(y)|^s}{|u_b(x+1)+(E-\eta_x)u_b(x)|^s}\rho(\eta_x)d\eta_x\leq C \frac{\left\|\left( \begin{array}{c} u_b(y)\\
u_b(y+1) \end{array} \right)\right\|^s}{\left\|\left( \begin{array}{c} u_b(x)\\
u_b(x+1) \end{array} \right)\right\|^s}.
\end{equation}
Expressed in terms of the discrete transfer matrices $T(x,y,E)$, the
right hand side is equal to\\
$C\|(u_b(y),u_b(y+1))^t\|^s/\|T(x,y,E)(u_b(y), u_b(y+1))^t\|^s$.
Thus the required bound follows from \eqref{eq:transferbd} and
Lemma~5.1 of \cite{CKM}, the discrete analogue of
Lemma~\ref{lem:furstenberg}.

In order to prove \eqref{eq:transferbd}, we first note that
$u_b(x)$, $u_b(x+1)$ as well as $u_b(y)$ are all independent of
$\eta_x$. With this in mind the proof of \eqref{eq:transferbd} is
naturally divided into two cases

{\bf Case I:} $u_b(x)=0$, in this case, the left hand side of
\eqref{eq:transferbd} is simply $\left|u_b(y)/u_b(x+1)\right|^s$
which is bounded above by $\left\|(u_b(y), u_b(y+1))^t\right\|^s /
\left\| (u_b(x), u_b(x+1))^t\right\|^s$.

{\bf Case II:} If $u_b(x)\neq 0$, let $M= \sup\{|E-\eta|: \eta\in
[\eta_{\rm min},\eta_{\rm max}], E\in I\}$. If $|u_b(x+1)/u_b(x)|>
2M$, then
\begin{eqnarray*}
\left|\frac{u_b(y)}{u_b(x)}\right|^s\int_{\eta_{\rm min}}^{\eta_{\rm max}}  \dfrac{\rho(\eta_x)}{|\frac{u_b(x+1)}{u_b(x)}+E-\eta_x|^s}\, d\eta_x & \leq & 2^s\|\rho\|_\infty (\eta_{\rm max}-\eta_{\rm min})\left|\frac{u_b(y)}{u_b(x+1)}\right|^s \\
& \leq & 2^s\|\rho\|_\infty (\eta_{\rm max}-\eta_{\rm min})\left(1+\frac{1}{4M^2}\right)^{s/2}\frac{\left\|\left(\begin{array}{c} u_b(y)\\
u_b(y+1) \end{array}\right)\right\|^s}{\left\| \left(\begin{array}{c} u_b(x)\\
u_b(x+1) \end{array} \right)\right\|^s}.
\end{eqnarray*}

  On the other hand if $|u_b(x+1)/u_b(x)|\leq 2M$, using that for any $\beta\in \Cmplx$ we have
\begin{equation} \label{eq:elembound}
\int_{\eta_{\rm min}}^{\eta_{\rm max}}
\dfrac{1}{|\beta-\eta_x|^s}\rho(\eta_x) \,d\eta_x\leq C_1(s,\rho),
\end{equation}
we see that
\begin{eqnarray*}
\left|\frac{u_b(y)}{u_b(x)}\right|^s\int_{\eta_{\rm min}}^{\eta_{\rm max}}  \dfrac{\rho(\eta_x)}{|\frac{u_b(x+1)}{u_b(x)}+E-\eta_x|^s} \, d\eta_x & \leq & C_1(s,\rho)\left(1+4M^2\right)^{s/2} \dfrac{|u_b(y)|^s}{\left\| \left(\begin{array}{c} u_b(x)\\
u_b(x+1) \end{array} \right)\right\|^s}
\\ & \leq & C(s,M,\rho) \frac{\left\|\left(\begin{array}{c} u_b(y)\\
u_b(y+1) \end{array}\right)\right\|^s}{\left\|\left(\begin{array}{c} u_b(x)\\
u_b(x+1) \end{array} \right)\right\|^s}
\end{eqnarray*}
 We have thus established \eqref{eq:transferbd}, which ends the proof.
\end{proof}

\section{Remarks} \label{sec:remarks}

(i) The proof of Lemma~\ref{lem:bounddis} works for
multi-dimensional discrete Anderson models without any changes.

(ii) With only minor changes the proofs of Lemma~\ref{lem:bounddis}
and Theorem~\ref{thm:1dfracdis} extend to complex energy. In
particular, this uses that the bound \eqref{eq:fracint} holds
uniformly in all dissipative matrices $B$ (as the matrices
$A_{\beta}$ are now dissipative) and that \eqref{eq:elembound}, the
scalar version of \eqref{eq:fracint}, holds uniformly in $\beta \in
\Cmplx$.

As a consequence, we see that the exponential decay bound
\eqref{eq:expodecdis} holds uniformly in $E\in \Cmplx$.

Working at complex energy our arguments in
Section~\ref{sec:discrete} may also be used to establish the
analogue of \eqref{eq:expodecdis} for infinite volume, i.e.\ to show
that
\[
\EE(|G(x,y;E+i\varepsilon)|^s) \le Ce^{-\eta|x-y|}
\]
holds uniformly in $E\in \Reals$, $\varepsilon\not= 0$, where
$G(x,y;z) = \langle e_x, (h-z)^{-1} e_y \rangle$. The only change is
that $u_b$ in \eqref{eq:greenformula} is replaced by $u_{\infty}$,
the unique solution (up to a scalar) of $-u(n-1)-u(n+1)+\eta_n
u(n)=(E+i\varepsilon)u(n)$ which is square-summable at $+\infty$.

(iii) While we expect that Theorem~\ref{thm:1dfrac} extends to
complex energy as well, we do not know how to get this with our
method of proof. The main problem here is that the Pr\"ufer
formalism strongly hinges on working with real-valued solutions. Due
to its usefulness in applications, it would be interesting to find a
different argument to allow for this extension.

\section*{Appendix: Basic facts} \label{sec:app}

In this section, we will collect some basic facts about Pr\"ufer
variables and two basic a-priori solution estimates which we use
repeatedly throughout the main text. A priori solution estimates
like Lemma~\ref{lem:solest} and Lemma~\ref{lem:L2b} are standard
tools in the theory of Sturm-Liouville operators.
Lemma~\ref{lem:lamderphi} as well as its Corollary~\ref{cor:Ederphi}
have been frequently used in connection with spectral averaging
techniques, e.g.\ \cite{Carmona/Lacroix}. We provide their proofs
merely to make the paper self-contained.

Throughout this appendix, with the exception of the last corollary,
the energy parameter $E$ will be absorbed in the potential term.

\begin{lem} \label{lem:solest}
For every $q\in L^1_{\rm loc}(\Reals)$, every interval $[c,d]$, and
every solution $u$ of $-u''+qu=0$ on $[c,d]$ one has that
\begin{eqnarray} \label{eq:solbd}
(|u(c)|^2 +|u'(c)|^2) \exp \left(- \int_c^d |1+q(x)|\, dx  \right) &
\leq & |u(d)| ^2 + |u'(d)|^2 \\ & \leq  & \left( |u(c)|^2 +
|u'(c)|^2 \right) \exp \left( \int_c^d |1+q(x)|\, dx  \right).
\nonumber
\end{eqnarray}
\end{lem}

\begin{proof} Setting $R(t) := |u(t)|^2+|u'(t)|^2$, one easily calculates that
\[
R'(t) \, = \,  2{\rm Re}\left[ \left( 1+q(t)\right) u(t)
\overline{u'(t)} \right],
\]
and hence
\begin{equation} \label{eq:dRbd}
| R'(t)| \, \leq \,  \left| 1 +q(t)\right| \, R(t).
\end{equation}
Since (\ref{eq:dRbd}) bounds the derivative of the logarithm of
$R(t)$, the lemma is proven.
\end{proof}

\begin{lem} \label{lem:L2b}
For any positive real numbers $\ell$ and $M$ there exists $C>0$ such
that
\begin{equation} \label{eq:intbd2}
\int_c^{c+\ell} |u(t)|^2 dt \geq C \left( |u(c)|^2 + | u'(c)|^2
\right)
\end{equation}
for every $c\in \Reals$, every $L^1_{\rm loc}$-function $q$ with
$\int_c^{c+\ell} |q(t)|\,dt \le M$, and any solution $u$ of
$-u''+qu=0$ on $[c,c+\ell]$.
\end{lem}

\begin{proof} First, we observe that, by rescaling, it is
sufficient to prove (\ref{eq:intbd2}) for real valued solutions with
$|u(c)|^2 + |u'(c)|^2 = 1$. By Lemma \ref{lem:solest}, there are
constants $0<C_1, C_2<\infty$, depending only on $\ell$ and $M$ for
which any real-valued solution of  $-u'' + qu = 0$ satisfies
\[
C_1 \,  \leq \, |u(x)|^2 + |u'(x)|^2 \, \leq \, C_2 \, ,
\]
for all $x \in [c,c+\ell]$; given the above mentioned normalization.
With $C_3 := (C_1/2)^{1/2}$ and $C_4 := (2C_2)^{1/2}$, we also have
that
\begin{equation} \label{eq:sum}
C_3 \,  \leq \, |u(x)| + |u'(x)| \, \leq \, C_4 \, .
\end{equation}

We now claim that for every $0< \alpha < \ell(2+ \ell)^{-1}$  exists
an $x_0( \alpha) = x_0 \in [c,c+\ell]$ for which
\begin{equation} \label{eq:uatx0}
|u(x_0) | \, \geq  \, \alpha \, C_3 \, .
\end{equation}
If, for such a fixed value of $\alpha$, this is not the case, then
for all $x \in [c,c+\ell]$,
\[
|u(x) | <  \alpha C_3,
\]
and from  (\ref{eq:sum}) we may also conclude that
\[
|u'(x)|  \, \geq  \, C_3 \,  - \, |u(x)| \, >  \, (1 \, - \, \alpha)
\, C_3 \,  > \, 0.
\]
Hence the derivative, $u'$, is strictly signed. With this we may
estimate,
\begin{eqnarray*}
2 \alpha C_3 \, > \, \left| u(c+\ell)-u(c) \right| & = &  \left| \int_c^{c+\ell} u'(x) \, dx \right|  \\
& = &   \int_c^{c+\ell} |u'(x)| \, dx  \\
 & > & (1 \, - \, \alpha) C_3 \ell.
\end{eqnarray*}
This contradicts the initial assumption on the range of $\alpha$,
and we have proven (\ref{eq:uatx0}).

The bound (\ref{eq:intbd2}) now follows as
\[
\left| u(x) - u(x_0) \right| \, \leq \, \int_{x_0}^x | u'(t)| \, dt
\, \leq \, C_4 \, |x-x_0|,
\]
implies that, in particular, $|u(x)| \geq \alpha C_3 /2$ for all $x
\in [c,c+\ell]$ for which $|x-x_0| \leq \alpha C_3 /(2 C_4)$.
\end{proof}

Our remaining results relate to Pr\"ufer variables. In general, for
any real potential $q\in L^1_{\rm loc}( \mathbb{R})$ and real
parameters $c$ and $\theta$ let $u_c$ be the solution of
\[
-u'' + q u = 0
\]
with $u_c(c)=\sin \theta$, $u_c'(c)=\cos \theta$. By regarding this
solution and its derivative in polar coordinates, we define the
Pr\"ufer amplitude $R_c(x)$ and the Pr\"ufer phase $\phi_c(x)$ by
writing
\begin{equation} \label{eq:pruferdef2}
u_c(x)= R_c(x) \sin \phi_c(x) \quad \mbox{and} \quad u_c'(x) =
R_c(x) \cos \phi_c(x).
\end{equation}
For uniqueness of the Pr\"ufer phase we declare $\phi_c(c)=\theta$
and require continuity of $\phi_c$ in $x$. In what follows the
initial phase $\theta$ will be fixed and we thus leave the
dependence of $u_c$, $R_c$ and $\phi_c$ implicit in our notation.

In the new variables $R$ and $\phi$ the second order equation
$-u''+qu=0$ becomes a system of two first order equations, where the
equation for $\phi$ is not coupled with $R$:
\begin{lem} \label{lem:xder} For fixed $c$ and $\theta$, one has
  that
\begin{equation} \label{eq:xderR}
(\ln R^2_c(x))'  \, = \, \left( 1 + q(x) \right) \, \sin \left( 2 \,
\phi_c(x) \right),
\end{equation}
and
\begin{equation} \label{eq:xderphi}
\phi_c'(x) \, = \, 1 \, - \, \left( 1 + q(x)\right) \, \sin^2 \left(
\phi_c(x) \right).
\end{equation}
\end{lem}
\begin{proof}
It is clear that $R^2_c = u^2 + (u')^2$, and (\ref{eq:xderR})
follows from a simple calculation. To see (\ref{eq:xderphi}),
observe the following two equations: $u' = R_c' \sin( \phi_c) + R_c
\cos( \phi_c) \phi_c'$ and $qu = u'' = R_c' \cos( \phi_c) - R_c
\sin( \phi_c) \phi_c'$. Solving for $\phi_c'$ yields
(\ref{eq:xderphi}).
\end{proof}

We have the following formula for the derivative of the Pr\"ufer
phase with respect to a coupling constant at a potential.

\begin{lem} \label{lem:lamderphi} Let $W$ and $V$ be real valued
  functions in $L^1_{\rm loc}( \mathbb{R})$. For real parameters $c$, $\theta$
  and $\lambda$, let $u_c$ be the solution of
\[
-u'' +  Wu + \lambda V u = 0
\]
normalized so that $u_c(c) = \sin( \theta)$ and $u_c'(c) = \cos(
\theta)$. Denoting the Pr\"ufer variables of $u_c$ by
$\phi_c(x,\lambda)$ and $R_c(x,\lambda)$, indicating their
dependence on the coupling constant $\lambda$, one has that
\begin{equation} \label{eq:lamder}
\frac{\partial}{\partial \lambda} \phi_c(x,\lambda) \, = \,
-R_c^{-2}(x,\lambda) \, \int_c^x V(t) \, u_c^2(t,\lambda) \, dt.
\end{equation}
\end{lem}

\begin{proof}
Using both (\ref{eq:xderR}) and (\ref{eq:xderphi}) from
Lemma~\ref{lem:xder} above, one finds that
\[
\frac{ \partial^2}{ \partial \lambda \partial x} \phi_c(x,\lambda) =
- V(x) \, \sin^2 \left( \phi_c(x, \lambda) \right) - \, \frac{
\partial}{ \partial x} \ln \left[ R_c^2(x, \lambda) \right] \,
\frac{ \partial}{
\partial \lambda} \phi_c(x, \lambda),
\]
This implies that
\begin{equation} \label{eq:xderfun2}
\frac{\partial}{\partial x} \left( R_c^2(x, \lambda) \,
  \frac{\partial}{\partial \lambda} \phi_c(x,\lambda) \right) \, = \,
-V(x) R_c^2(x, \lambda) \, \sin^2 \left( \phi_c(x, \lambda) \right)
\, = \, =-V(x)u_c^2(x,\lambda),
\end{equation}
for almost every pair $(x,\lambda)$. Since $\frac{\partial}{\partial
\lambda} \phi_c(c,\lambda) \, = \, 0$, (\ref{eq:lamder}) follows
immediately from (\ref{eq:xderfun2}).
\end{proof}

As a special case one finds the energy derivative of the Pr\"ufer
phase.

\begin{cor} \label{cor:Ederphi}
Let $u$ be the solution of $-u'' +  Wu = Eu$ normalized so that
$u(c) = \sin( \theta)$ and $u'(c) = \cos( \theta)$, and let
$\phi_c(x,E)$ and $R_c(x,E)$ be the corresponding Pr\"ufer
variables. Then
\begin{equation} \label{eq:Ederphi}
\frac{ \partial}{ \partial E} \phi_c(x,E) \, = \, R_c^{-2}(x,E) \,
\int_c^x u^2(t) \, dt.
\end{equation}
\end{cor}
\begin{proof}
This follows from Lemma~\ref{lem:lamderphi} by setting $V$ constant
to $-1$.
\end{proof}

\noindent {\bf Acknowledgements:}
 A part of this work was supported
by the National Science Foundation, e.g., R.S. under Grant
\#DMS-0757424 and G.S. under Grant DMS-0653374.

\end{document}